\documentstyle[12pt]{article}
\begin{document}

\def\lta{\;\raisebox{-.5ex}{\rlap{$\sim$}} \raisebox{.5ex}{$<$}\;}
\def\gta{\;\raisebox{-.5ex}{\rlap{$\sim$}} \raisebox{.5ex}{$>$}\;}
\def\grle{\;\raisebox{-.5ex}{\rlap{$<$}}    \raisebox{.5ex}{$>$}\;}
\def\legr{\;\raisebox{-.5ex}{\rlap{$>$}}    \raisebox{.5ex}{$<$}\;}

\newcommand{\ra}{\rightarrow}
\newcommand{\permille}{$^0 \!\!\!\: / \! _{00}$}
\newcommand{\dd}{{\rm d}}
\newcommand{\oal}{{\cal O}(\alpha)}%
\newcommand{\su}{$ SU(2) \times U(1)\,$}
 
\newcommand{\eps}{\epsilon}
\newcommand{\mw}{M_{W}}
\newcommand{\mww}{M_{W}^{2}}
\newcommand{\mbb}{m_{b \bar b}}
\newcommand{\mcc}{m_{c \bar c}}
\newcommand{\mbc}{m_{b\bar b(c \bar c)}}
\newcommand{\mh}{m_{H}}
\newcommand{\mhh}{m_{H}^2}
\newcommand{\mz}{M_{Z}}
\newcommand{\mzz}{M_{Z}^{2}}

\newcommand{\lra}{\leftrightarrow}
\newcommand{\tr}{{\rm Tr}}
 
\newcommand{\ie}{{\em i.e.}}
\newcommand{\cm}{{{\cal M}}}
\newcommand{\cl}{{{\cal L}}}
\def\Ww{{\mbox{\boldmath $W$}}}  
\def\B{{\mbox{\boldmath $B$}}}         
\def\nn{\noindent}

\newcommand{\sinsq}{\sin^2\theta}
\newcommand{\cossq}{\cos^2\theta}

\newcommand{\epem}{$e^{+} e^{-}\;$}
\newcommand{\epemt}{e^{+} e^{-}\;}
\newcommand{\eeah}{$e^{+} e^{-} \ra H \gamma \;$}
\newcommand{\eahnw}{$e\gamma \ra H \nu_e W$}

\newcommand{\thebb}{\theta_{b-beam}}
\newcommand{\thebc}{\theta_{b(c)}}
\newcommand{\pte}{p^e_T}
\newcommand{\ptH}{p^H_T}
\newcommand{\gag}{$\gamma \gamma$ }
\newcommand{\gam}{\gamma \gamma }

\newcommand{\aatoh}{$\gamma \gamma \ra H \;$}
\newcommand{\egam}{$e \gamma \;$}
\newcommand{\eat}{e \gamma \;}
\newcommand{\eaeh}{$e \gamma \ra e H\;$}
\newcommand{\eaehb}{$e \gamma \ra e H \ra e (b \bar b)\;$}
\newcommand{\egebb}{$e \gamma (g) \ra e b \bar b\;$}
\newcommand{\egecc}{$e \gamma (g) \ra e c \bar c\;$}
\newcommand{\egebc}{$e \gamma (g) \ra e b \bar b(e c \bar c)\;$}
\newcommand{\eaebb}{$e \gamma \ra e b \bar b\;$}
\newcommand{\eaecc}{$e \gamma \ra e c \bar c\;$}
\newcommand{\aah}{$\gamma \gamma H\;$}
\newcommand{\zah}{$Z \gamma H\;$}
\newcommand{\pe}{P_e}
\newcommand{\pg}{P_{\gamma}}
\newcommand{\delbb}{\Delta m_{b \bar b}}
\newcommand{\delbc}{\Delta m_{b \bar b(c\bar c)}}


\title{\Large \bf  Testing the $Z\gamma H$ vertex at future linear
       colliders for intermediate Higgs masses\footnote{Contribution to the
       Proceedings of the Int. Europhysics Conference on High Energy Physics
       (Jerusalem, 19-26 August, 1997)}
      }

\author{E.~Gabrielli$^a$, V.A.~Ilyin$^b$ and B.~Mele$^c$ \\ \\
 {\small \it $^a$ University of Notre Dame, IN, USA} \\
 {\small \it $^b$ Institute of Nuclear Physics, 
                  Moscow State University, Russia} \\
 {\small \it $^c$ INFN, Sezione di Roma 1 and Rome University "La Sapienza", 
                  Italy}
        }
\date{}

\maketitle

\begin{abstract}
Higgs production in $e\gamma$ collisions, through the one-loop reaction \eaeh
at large $p_T$, can provide a precise determination of the $Z\gamma H$ vertex. 
\end{abstract}

Among other couplings, the interactions of the Higgs scalar with $\gamma$ and
$Z$ are particularly interesting, since they depend on the relation between the
spontaneous symmetry breaking mechanism and the electroweak mixing of the two
gauge groups $SU(2)$ and $U(1)$. In this respect, three vertices can be
studied:  $ZZH$, \aah and \zah. While in the SM the $ZZH$ vertex stands at the
tree level, the other two contribute only at one-loop. This means that the \aah
and \zah couplings can be sensitive to the contributions of new particles
circulating in the loop. For the Higgs masses discussed here, $\mh\lta 140$
GeV, a measurement of the \aah coupling should be possible by the determination
of the BR for the decay $H\ra\gamma\gamma$, e.g. in the LHC Higgs discovery
channel, $gg\to H\to\gamma\gamma$, or in $\gamma\gamma\to H$ at future
$\gamma\gamma$ linear colliders. A chance of measuring the \zah vertex is given
by collision processes, e.g. in $e^+e^-\to\gamma H\,,Z H$. However, in the $ZH$
channel the \zah vertex contributes to the one-loop corrections, thus implying
a large tree-level background. The reaction $e^+e^-\to \gamma H$ has been
extensively studied \cite{ee-aH}. Unfortunately, it suffers from small rates,
$\approx 0.05\div 0.001$ fb at $\sqrt{s}\sim 500\div 1500$ GeV, and (as we
estimated) the main background $e^+e^-\to\gamma b\bar b$ process has large
cross sections: $\approx 4\div 0.8$ fb for $m_{b\bar b}=100\div 140$ GeV, at
$\sqrt{s}\sim 500\div 1500$GeV, assuming reasonable kinematical cuts. Recently,
the one-loop process \eaeh was analysed in details \cite{noii}. The total rate
for this reaction is rather high,  $>1$ fb for $\mh<$ 400 GeV. The main
strategy to enhance the \zah vertex effects consists in requiring a final
electron tagged at large angle. E.g., for $\pte>100$GeV, \zah is about 60\% of
the generally dominant \aah contribution.  The main irreducible background
comes from \eaebb. A further background is the charm production through \eaecc,
when the $c$ quarks are misidentified into $b$'s.  At $\sqrt{s}= 500$ GeV the
cut $\thebc>18^\circ$ (between each $b(c)$ quark and the beams) makes the
background comparable to the signal \cite{noii}. Resolved \egebc production,
where the photon interacts via its gluonic content, could also contribute but,
as we found, it is quite small. We studied polarization effects and found they
are rather strong.  E.g., for right handed electrons there is a strong
destructive interference between the terms \aah and \zah.

Now we discuss the prospects of the \eaeh reaction in setting experi\-men\-tal
bounds on a possible anomalous \zah coupling. We assume, that the ano\-ma\-lous
\aah contributions have been well tested in some other experiment (e.g.,
through $\gamma\gamma\to H$). Then, one would like to get limitations just on
the anomalous \zah contributions. Anomalous CP-even and CP-odd operators 
contributing to \eaeh \cite{anomOP} are:
$$ {\cal O}_{UW;UB} \,=\, \left( \frac{|\Phi|^2}{v^2}-\frac{1}{2}\right)
     \left\{W W; B B \right\}\;,
 \qquad
 \bar {\cal O}_{UW;UB} \,=\, \frac{|\Phi|^2}{v^2}  
     \left\{ W \tilde W;
                  B  \tilde B \right\}\;,
$$
where
$ {\cal L}^{eff} = d\cdot {\cal O}_{UW} + d_B\cdot {\cal O}_{UB} +
     \bar d\cdot \bar {\cal O}_{UW} 
                        + \bar d_B\cdot \bar {\cal O}_{UB}$.
The corresponding $Z\gamma H$ anomalous terms in the helicity  amplitudes of
$e\gamma\to eH$ are
$$ 
  \frac{4\pi\alpha}{M_Z(M_Z^2-t)}  \sqrt{-\frac{t}{2}}
   \left\{ d_{\gamma Z} [(u-s)-\sigma\lambda (u+s)] 
   - i{\bar d_{\gamma Z}} [\lambda (u-s)+\sigma (u+s)] \right\},
$$
where $s$, $t$ and $u$ are the  Mandelstam kinematical variables, $\sigma/2$
and $\lambda$ are the electron and photon  helicities, and $d_{\gamma
Z}=d-d_B$, $\bar d_{\gamma Z}=\bar d-\bar d_B$.

At $\sqrt{s}=500$ GeV, for $m_H=120$ GeV, one can then constrain the CP-even 
coupling in the following way: $-0.0025<d_{\gamma Z}<0.004$ in the unpolarized
case, $|d_{\gamma Z}|<0.0015$ for left-handed and $-0.007 <d_{\gamma Z}<0.004$
for right-handed electrons. The corresponding bounds on the CP-odd coupling
depends only slightly on the electron polarization, and  are $|\bar d_{\gamma
Z}|\lta 0.006$. Here we have taken into account the  contributions for
background from $e\gamma\to eb\bar b(ec\bar c)$, assuming 10\% of the $c/b$
misidentifying, and from the resolved photons. The cuts $\thebc>18^\circ$,
$\pte>100$ GeV and  $|\mbc-\mh|<3$ GeV are applied. The bounds presented 
have been computed by using the requirement that no deviation from the SM cross
section is observed at the 95\% CL, with an integrated luminosity 100
fb$^{-1}$. If the anomalous terms appear as contributions of new particles in
the \zah loop with the mass $M_{new}$, then one gets $d_{\gamma Z},\bar
d_{\gamma Z}\sim (v/M_{new})^2$. By using this relation, one obtains the bounds
$M_{new}\gta 6.2$ TeV in the CP-even case and $M_{new}\gta 3.5$ TeV in the
CP-odd case. All the results presented here were obtained with the help of the
CompHEP package \cite{comp}.

\end{document}